\documentclass[reprint,showpacs,preprintnumbers,amsmath,amssymb,floatfix]{revtex4-1}
\usepackage{color}
\usepackage{graphicx}% Include figure files
\usepackage{dcolumn}% Align table columns on decimal point
\usepackage{bm}% bold math
\usepackage{amstext}
\usepackage{amsmath}

\begin{document}

\title{Revisiting the thermal relaxation of neutron stars}

\author{Thiago Sales}
\affiliation{Instituto de F\'isica, Universidade Federal Fluminense 20420, Niter\'oi, RJ, Brazil}

\author{Odilon Louren\c{c}o}
\affiliation{\mbox{Departamento de F\'isica, Instituto Tecnol\'ogico de Aeron\'autica, DCTA, 12228-900, S\~ao Jos\'e dos Campos, SP, Brazil}}

\author{Mariana Dutra}
\affiliation{\mbox{Departamento de F\'isica, Instituto Tecnol\'ogico de Aeron\'autica, DCTA, 12228-900, S\~ao Jos\'e dos Campos, SP, Brazil}}

\author{Rodrigo Negreiros}
\affiliation{Instituto de F\'isica, Universidade Federal Fluminense 20420, Niter\'oi, RJ, Brazil}

\begin{abstract}
In this work we revisit the thermal relaxation process for neutron stars. Such process is associated with the thermal coupling between the core and the crust of neutron stars. The thermal relaxation, which takes place at around 10 -- 100 years, is manifested as a sudden drop of the star's surface temperature. Such drop is smooth for slow cooling objects and very sharp for those with fast cooling. In our study we focus particularly on the cooling of neutron stars whose mass is slightly greater than the value above which the direct Urca (DU) process sets in. Considering different mechanisms for neutrino production in each region of the star, and working with  equations of state with different  properties, we solve the thermal evolution equation and calculate the thermal relaxation time for ample range of neutron star masses. By performing a comprehensive  study of neutron stars just above the onset of the direct Urca process we show that stars under these conditions exhibit a peculiar thermal relaxation behavior. We demonstrate that such stars exhibit an abnormally late relaxation time, characterized by a second drop of its surface temperature taking place a later ages. We qualify such behavior by showing that it is associated with limited spatial distribution of the DU process is such stars. We show that as the star's mass increase, the DU region also grows and the start exhibits the expected behavior of fast cooling stars. Finally we show that one can expect high relaxation times for stars in which the DU process takes place in a radius not larger than 3 km.

\end{abstract}

\maketitle
\section{Introduction}
The cooling of neutron stars has demonstrated to be a fantastic way of probing the interior of these objects. Many works were dedicated to investigating several aspects of this rich and complicated phenomena \cite{TSURUTA1965,Maxwell1979,Horvath1991,1996NuPhA.605..531S,2004ApJS..155..623P,Grigorian2005,Page2006,Page2011a,2013PhLB..718.1176N,2017A&A...603A..44N,2018ApJ...863..104N}. Our current understanding of the thermal evolution of these objects tell us that they cool down mainly due to two mechanisms: neutrino emission from their interior and photon emission from the surface \citep[for a comprehensive review see][]{Yakovlev2004,2004ApJS..155..623P,Page2009}. Initially the neutrino emission from the interior dominates the cooling. After this neutrino dominated era, when the interiors are cool enough so that neutrino emission becomes less relevant, the cooling is driven by photon emission from its surface. Furthermore, the significant differences between the structure of the star's core and crust (the former composed of a degenerate interacting gas whereas the latter is mostly crystalline)  lead to a thermal decoupling between them. Due to stronger emission, the core acts as a heat sink,  absorbing part of the crust's heat (while the other part is radiated away at the surface and by crustal neutrino emission). Eventually the core and the crust become thermally coupled, a process that is signaled by a drop of the surface temperature of the star \cite{Lattimer1994,Potekhin1997,Gnedin2001}. The drop in temperature is more or less accentuated according to how strong the neutrino emission in the core is. Usually, stars in which the powerful direct Urca (DU) process is taking place will exhibit a significant and sharp temperature drop, whereas stars without the DU will have a much smoother thermal evolution ~\cite{Lattimer1991,Yakovlev2001}. It is this sudden surface temperature change that is used to define the thermal relaxation time.

Previous works \cite{Lattimer1994,Gnedin2001} have found that the thermal relaxation time depends  on microscopic properties of the star, such as thermal conductivity and specific heat. It was also found that it  depends on macroscopic properties such as stellar radius and crust thickness. In this work we will revisit the thermal processes that lead to the thermal relaxation. We will show that while our results agree with the previous studies, we identified a new behavior: stars that exhibit an abnormally high thermal relaxation time. Our work identifies that this transient behavior is typical of stars just above the onset of the DU process. We will show that this is associated with the small regions in which the DU process is active is such stars (as opposed to larger regions in more massive stars, or complete absence for lower mass objects). We will see that for the stars in which the DU is not pervasive its thermal relaxation can take much longer - and that these stars exhibit thermal behavior typical of stars with and without DU process. 

In order to perform this study we make use of several microscopic models that have been extensively used and tested for modelling neutron stars \cite{dutra2014,PhysRevC.99.045202}. From the pool of models studied in \cite{dutra2014,PhysRevC.99.045202} we have chosen four: BSR8, BSR9, G2* and IU-FSU -- all of them allow for the DU process to set in at reasonable neutron star masses (between 1.0 and 2.0 solar masses). Furthermore, all of them present microscopic properties different enough to allow us to conclude that the results we find are most likely general. As we will see, the same behavior is shared among all models studied, with the only difference being the star's mass at which the DU process becomes available. 

This paper is divided as follow: in section 2 we discuss the microscopic models used, section 3 is devoted to the review of the thermal evolution of neutron stars, section 4 contains our results for the relaxation time of neutron stars at the onset of the direct Urca process, and in section 5 we present our conclusions.

\section{Microscopic Models} \label{Micro}

The Quantum Hadrodynamics (QHD) is a powerful tool used to build models that represent the strongly interacting matter with hadrons being the mainly degrees of freedom. The first model constructed from such an approach was proposed by Walecka~\cite{walecka1,walecka2} with the two free parameters fixed to reproduce the nuclear matter energy per particle as a function of the density, with a minimum of $B_0=-15.75$~MeV at the saturation density $\rho_0=0.19$~fm$^{-3}$. However, the model also presents bad results for the effective mass ratio and incompressibilty, both at $\rho=\rho_0$, namely, $M^*_0/M_{\mbox{\tiny n}}=0.56$ ($M_{\mbox{\tiny nuc}}$ is the nucleon rest mass) and $K_0=540$~MeV, respectively. Over the years, many other improved versions of this model were proposed in which $M^*_0/M_{\mbox{\tiny n}}$ and $K_0$ are fixed to more compatible values with experimental/theoretical predictions. Furthermore, other bulk parameters at the saturation density are also used in order to constrain the free coupling constants of these microscopic models. 

Here, we investigate the cooling process of neutron stars based on parametrizations of a general model described by the following Lagrangian density~\cite{dutra2014,LI2008113},
\begin{eqnarray}
\mathcal{L} &=& \bar{\psi}\left(i \gamma^{\mu} \partial_{\mu}-M_{\mbox{\tiny n}}\right) \psi+g_{\sigma} \sigma \bar{\psi} \psi-g_{\omega} \bar{\psi} \gamma^{\mu} \omega_{\mu} \psi + \frac{m_{\rho}^{2}}{2}  \vec{\rho}_{\mu} \vec{\rho}^{\mu}
\nonumber\\ 
&-&\frac{g_{\rho}}{2} \bar{\psi} \gamma^{\mu} \vec{\rho}_{\mu} \vec{\tau} \psi
+\frac{1}{2}\left(\partial^{\mu} \sigma \partial_{\mu} \sigma-m_{\sigma}^{2} \sigma^{2}\right)-\frac{A}{3} \sigma^{3}-\frac{B}{4} \sigma^{4} \nonumber\\
&-&\frac{1}{4} F^{\mu \nu} F_{\mu \nu}+\frac{1}{2} m_{\omega}^{2} \omega_{\mu} \omega^{\mu}+\frac{C}{4}\left(g_{\omega}^{2} \omega_{\mu} \omega^{\mu}\right)^{2} -\frac{1}{4} \vec{B}^{\mu \nu} \vec{B}_{\mu \nu}
\nonumber\\
&+& g_\sigma g_\omega^2\sigma\omega_\mu\omega^\mu\left(\alpha_1+\frac{1}{2}{\alpha'_1}g_\sigma\sigma\right)
+\frac{1}{2} \alpha_3' g_{\omega}^{2} g_{\rho}^{2} \omega_{\mu} \omega^{\mu} \vec{\rho}_{\mu} \vec{\rho}^{\mu}
\nonumber\\
&+& g_\sigma g_\rho^2\sigma\vec{\rho}_\mu\vec{\rho}^\mu  \left(\alpha_2+\frac{1}{2}{\alpha'_2}g_\sigma\sigma\right),
\label{lagrangian} 
\end{eqnarray}
in which $F_{\mu \nu}=\partial_{\mu} \omega_{\nu}-\partial_{\nu} \omega_{\mu}$ and $\vec{B}_{\mu \nu}=\partial_{\mu} \vec{\rho}_{\nu}-\partial_{\nu} \vec{\rho}_{\mu}$. $\psi$ is the nucleon field and $\sigma$, $\omega_\mu$ and $\vec{\rho}_{\mu}$ represent the fields of the mesons $\sigma$, $\omega$ and $\rho$, respectively. The mean-field approximation is used in order to solve the equations of motion for the fields. This procedure, along with the energy-momentum tensor, $T_{\mu \nu}$, allows the construction of all thermodynamics of the system since the energy density and pressure are given by $\mathcal{E}=\langle T_{00}\rangle$ and $P=\langle T_{ii}\rangle /3$, respectively. These equations of state are evaluated, as a function of the density, by taking into account the auto-consistency of the field equations and the definition of the effective nucleon mass given by $M^*= M_{\mbox{\tiny n}}-g_\sigma\sigma$. More details related to the calculations of these quantities can be found in Refs.~\cite{dutra2014,LI2008113} and references therein.

In order to study neutron stars and their thermal evolution, it is needed to construct  stellar matter by imposing charge neutrality and $\beta$-equilibrium. This leads to the following conditions upon chemical potentials and densities: $\mu_n - \mu_p = \mu_e=\mu_\mu$ and $\rho_p - \rho_e = \rho_\mu$, where \mbox{$\rho_l=[(\mu_l^2 - m_l^2)^{3/2}]/(3\pi^2)$}, for $l=e, \mu$, and $\mu_e=(3\pi^2\rho_e)^{1/3}$. The total energy density and pressure of $\beta$-equilibrated stellar matter is then given by $\varepsilon = \mathcal{E} + \mathcal{E}_e + \mathcal{E}_\mu$ and $p = P + P_e + P_\mu$, respectively. The chemical potentials and densities of protons, neutrons, electrons and muons are given, 
respectively, by $\mu_p$, $\mu_n$, $\mu_e$, $\mu_\mu$, and $\rho_p$, $\rho_n$, $\rho_e$, $\rho_\mu$, with $y=\rho_p/\rho=\rho_p/(\rho_p+\rho_n)$. Some neutron star properties, such as the mass-radius profile, are obtained through the solution of the Tolman-Oppenheimer-Volkoff~(TOV) equations~\cite{tov39,tov39a} given by $dp(r)/dr=-[\varepsilon(r) + p(r)][m(r) + 4\pi r^3p(r)]/r^2f(r)$ and $dm(r)/dr=4\pi r^2\varepsilon(r)$, where $f(r)=1-2m(r)/r$. 

We choose to study parametrizations of the relativistic mean-field (RMF) model described by Eq.~(\ref{lagrangian}) that lead to the onset of the DU process \cite{Lattimer1991,Yakovlev2001}  at densities associated with a relatively wide range of stellar masses (see next section for more details), namely, BSR8~\cite{bsr89}, BSR9~\cite{bsr89}, G2*~\cite{g2s} and \mbox{IU-FSU}~\cite{iufsu}. Their main bulk properties (at the saturation density) are listed in Table~\ref{bulk}.
\begin{table}[!htb]
\centering
\caption{Bulk parameters of the RMF parametrizations used in the study of the neutron star cooling process.}
%\resizebox{\columnwidth}{!}{
\begin{tabular}{lrrrr}
\hline
Quantity                       & BSR8     & BSR9     & G2*        & IU-FSU \\
\hline
$\rho_0$ (fm$^{-3}$)           & $0.147$  & $0.147$  &  $0.154$   & $0.155$ \\
$B_0$ (MeV)                    & $-16.04$ & $-16.07$ &  $-16.07$  & $-16.40$ \\
$K_0$ (MeV)                    & $230.95$ & $232.50$ &  $214.77$  & $231.33$ \\
$M^*_0/M_{\mbox{\tiny n}}$     & $0.61$   & $0.60$   &  $0.66$    & $0.61$ \\
$J$ (MeV)                      & $31.08$  & $31.61$  &  $30.39$   & $31.30$ \\
$L_0$ (MeV)                    & $60.25$  & $63.89$  &  $69.68$   & $47.21$ \\
$K^0_{\mbox{\tiny sym}}$ (MeV) & $-0.74$  & $-11.32$ &  $-21.93$  & $28.53$ \\
\hline
\end{tabular}
%}
\label{bulk}
\end{table}
The isovector bulk parameters shown in that table are $J=\mathcal{S}(\rho_0)$ (symmetry energy at $\rho_0$), $L_0=L(\rho_0)$ (symmetry energy slope at $\rho_0$) and $K^0_{\mbox{\tiny sym}}=K_{\mbox{\tiny sym}}(\rho_0)$ (symmetry energy curvature at $\rho_0$), with $\mathcal{S}(\rho)=(1/8)(\partial^{2}E/\partial y^2)|_{y=1/2}$, $L_0=3\rho_0(\partial\mathcal{S}/\partial\rho)_{\rho_0}$, $K^0_{\mbox{\tiny sym}}=9\rho_0^2(\partial^2\mathcal{S}/\partial\rho^2)_{\rho_0}$ and $E(\rho)=\mathcal{E}/\rho$. These specific parametrizations were selected out of $35$ other ones shown to be consistent with constraints related to nuclear matter, pure neutron matter, symmetry energy, and its derivatives, in an analysis that investigated a larger set of $263$ RMF parametrizations~\cite{dutra2014}.

\section{Cooling of Neutron Stars}

The cooling of neutron stars is driven by the emission of neutrinos and photons, the former being emitted from the stellar core, and the latter from the surface. The thermal evolution equations for a spherically symmetric, relativistic star - with geometric unit system ($G = c = 1$) are given by,
\begin{eqnarray}
  \frac{ \partial (l e^{2\Phi})}{\partial m}& = 
  &-\frac{1}{\varepsilon \sqrt{1 - 2m/r}} \left( \epsilon_\nu 
    e^{2\Phi} + c_v \frac{\partial (T e^\Phi) }{\partial t} \right) \, , 
  \label{coeq1}  \\
  \frac{\partial (T e^\Phi)}{\partial m} &=& - 
  \frac{(l e^{\Phi})}{16 \pi^2 r^4 \kappa \varepsilon \sqrt{1 - 2m/r}} 
  \label{coeq2} 
  \, .
\end{eqnarray}

Details of the derivation of such equations can be found in references \citep{Page2006,1999Weber..book,1996NuPhA.605..531S}. One must also note that the cooling of neutron stars strongly depends on both micro and macroscopic properties of the star, which makes thermal evolution studies a fantastic way of probing compact star properties.
Quantities that are of extreme importance  to calculate the cooling are the neutrino emissivity  ($\epsilon_\nu(r,T)$),  thermal conductivity ($\kappa(r,T)$), and specific heat ($c_v(r,T)$), all of which depend on microscopic information of the underlying model. In addition to that, macroscopic properties such as radial distance ($r$), mass ($m(r)$), curvature ($\phi(r)$), temperature $T(r,t)$, luminosity ($l(r,t)$) are also needed for the solution of eqs.~(\ref{coeq1}) and (\ref{coeq2}).

The boundary conditions for the solution of eqs.~(\ref{coeq1}) and (\ref{coeq2}) are given  by a vanishing heat flow at the star's center $L(r=0) = 0$ and by the relationship between the surface luminosity and the mantle temperature - this last condition depends on the surface properties of the star and its composition, and is discussed in details in \cite{Gudmundsson1982,Gudmundsson1983}.

This study takes into account all neutrino emission processes allowed to happen in accordance with our current understanding - for a detailed review of such processes we direct the reader to references  \cite{Yakovlev2000,Yakovlev2004} . 

Notice that we intentionally do not consider pairing among the star's constituents. This is a conscious decision, as not to cloud the object of study, namely the thermal relaxation time. Evidently we are not advocating for the absence of pairing in neutron stars and refer the reader to several papers on the subject \cite{1996NuPhA.605..531S,2004ApJS..155..623P,Yakovlev2001,Page2011a}. In this work, however, we study the thermal evolution of objects without pairing, as to properly quantify and qualify the relaxation time of these stars. Evidently this work should be augmented with the inclusion of pairing - which is currently underway.

\subsection{Thermal Relaxation }
The thermal relaxation of a neutron star is characterized by the time it takes for the core and the crust to become thermally coupled. As explained in refs.~\cite{Lattimer1994,Gnedin2001}, due to the significantly different structures between the neutron star core and its crust (the latter composed roughly of a degenerate gas and the former of a crystalline structure) their thermal conductivity and specific heat are drastically different. Furthermore, there is strong emission of neutrinos in the core (the actual strength of the neutrino emission will depend on the presence or not of the direct Urca process, as we will discuss below). These two factors lead to the formation of cold front at the core that can be imagined to ``propagate'' towards the surface. Once it emerges, the surface temperature of the star exhibits a sudden drop (more or less accentuated according to the presence or absence of the DU process) - signaling the thermal coupling between the core and crust. Such relaxation times are typically $t_w \sim 10 - 100$ years - depending on stellar properties. In this work we follow the definition of \cite{Gnedin2001} and define the relaxation time as
\begin{equation}
    t_w  = \max \left| \frac{d\ln(T_s)}{d(\ln(t))} \right|. \label{tw}
\end{equation}

The authors of \cite{Lattimer1994,Gnedin2001} have found that the relaxation time can generally  be written as
\begin{equation}
    t_w \approx \alpha t_1, \label{tw1}
\end{equation}
where $t_1$ is a normalized relaxation time that depends solely on the composition of the star. $\alpha$ is given by
\begin{equation}
    \alpha = \left(\frac{\Delta R_{\text{crust}}}{1\text{km}}\right)^2 \left( 1- 2M/R\right)^{-3/2},
\end{equation}
where  $\Delta R_{\text{crust}}$ is the crust thickness, and $M$ and $R$ the stellar mass and radius, respectively. 

Eq.~(\ref{tw})  shows a linear dependence between the relaxation time and the quantity $\alpha$, which in turn strongly depends on macroscopic properties such as crust thickness, mass and radius. The normalized time $t_1$ is a propotionality constant that depends on the microscopic properties, such as specific heat and thermal conductivity of the underlying model \cite{Gnedin2001}. 

These results indicate that stars with higher mass, which are associated with thinner crusts and more intense neutrino emissions from their core, have a smaller relaxation time than their low mass counterparts - which is indeed the case as noted in \cite{Gnedin2001}. One notes a substantial difference between stars that exhibit fast cooling (generally higher mass objects with the presence of powerful neutrino emission processes such as the DU) and slow cooling stars (objects with lower masses). It was found that they both obey relation (\ref{tw}) but have different values for the coefficient ($t_1$) - which is understandable given that such coefficient is associated with microscopic properties of the star. In this work we investigate stars at the transition between slow and fast cooling regimes. We will see that such transition is generally non-linear, with the onset of a fast cooling process such as the DU giving rise to a substantial change in the thermal properties. We will show in the next section that in this transition regime stars exhibit  longer thermal relaxation times than their counterparts either in the slow or fast cooling regimes. 

\section{Relaxation time at the onset of the DU process}

Before we can devote our attention to the thermal behavior of stars near the onset of the DU process we will discuss the general cooling properties of the stars described by the different microscopic models described in section \ref{Micro}.
By varying the central density we calculate a family of stars whose cooling can then be calculated. Furthermore, we also identified the stars at which the DU process becomes active. We recall that the DU process can only take place if the triangle inequality $k_{fn} \leqslant k_{fp} + k_{fe}$ is satisfied. This usually translate to a proton fraction $\sim 11 - 15\%$\cite{Lattimer1991,Page2006}. The properties of the stars at the onset of DU process for the models studied are shown in Table~\ref{DU_threshold}. 

\begin{table}[!htb]
\centering
\caption{Stellar central energy density $\rho_{DU}$, stellar mass $M_{DU}$ and proton fraction $Y_{DU}$ above which the DU process is active inside the star for all EoS's studied herein.}
%\resizebox{\columnwidth}{!}{
\begin{tabular}{lccc}
\hline
Model           &$\rho_{DU}(fm^{-3})$    & $M_{DU}/M_{\odot}$   &$Y_{DU}$\\
\hline
BSR8          &$0.405$  & $1.41$
&$0.135$\\
BSR9          &$0.385$  & $1.31$
&$0.135$\\
G2*           &$0.390$  & $1.19$
&$0.135$\\
IU-FSU        &$0.614$  & $1.77$
&$0.138$\\
\hline
\end{tabular}
%}
\label{DU_threshold}
\end{table}

\begin{figure}[h!]
\centering
\includegraphics[width = 7 cm, height = 5 cm]{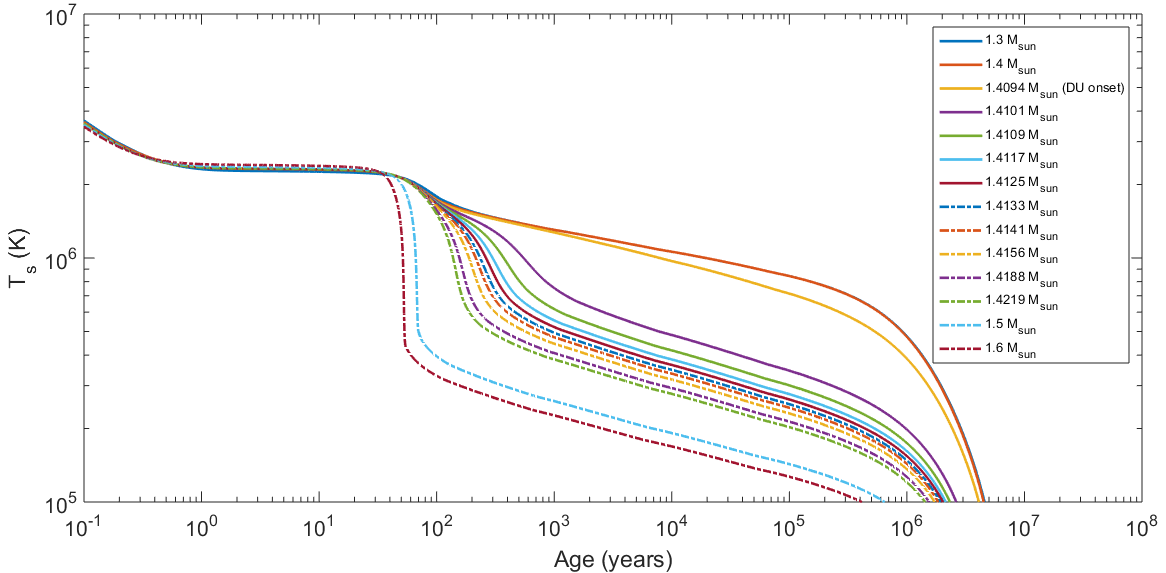}
\caption{Surface temperature as a function of age for stars of different masses under the microscopic model BSR8.}
\label{bsr8cool}
\end{figure}

\begin{figure}[h!]
\centering
\includegraphics[width = 7 cm, height = 5 cm]{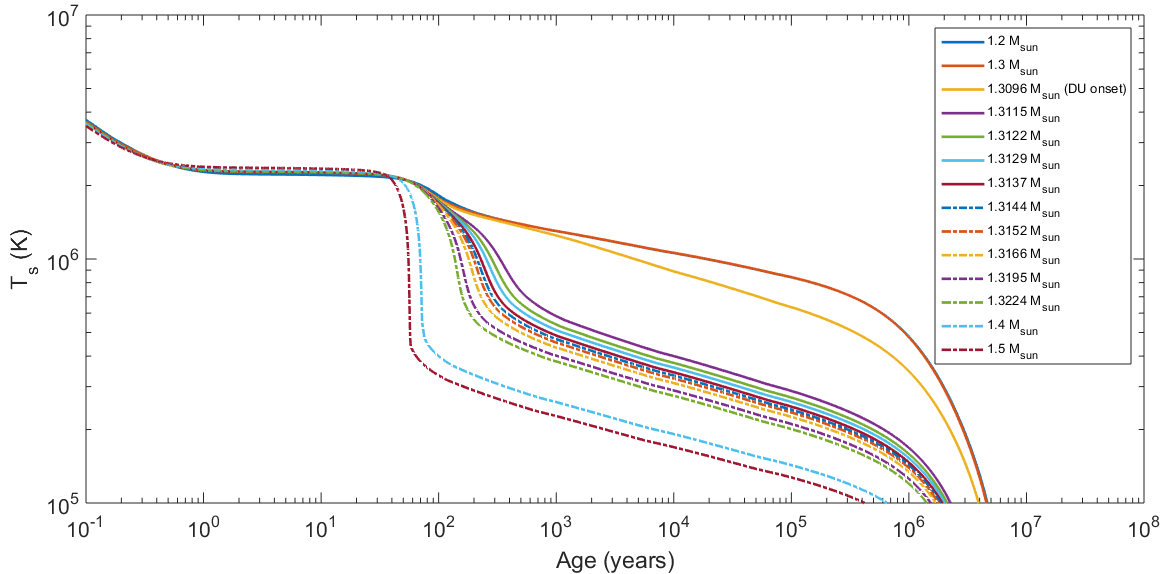}
\caption{Same as Fig.~\ref{bsr8cool}, but for model BSR9.}
\label{bsr9cool}
\end{figure}

\begin{figure}[h!]
\centering
\includegraphics[width = 7 cm, height = 5 cm]{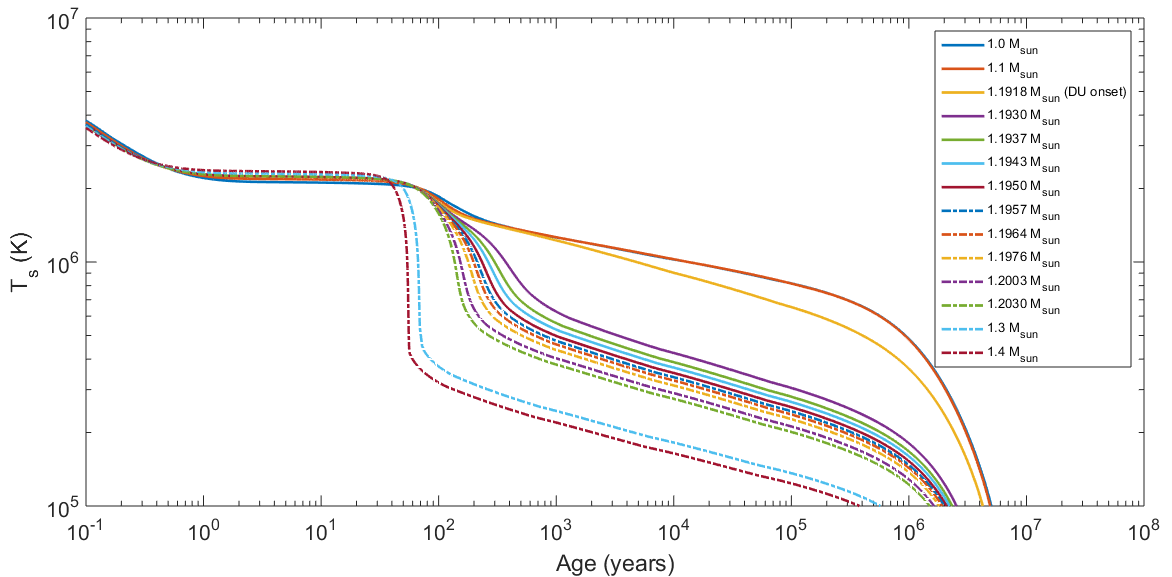}
\caption{Same as Fig.~\ref{bsr8cool}, but for model G2*.}
\label{g2starcool}
\end{figure}

\begin{figure}[h!]
\centering
\includegraphics[width = 7 cm, height = 5 cm]{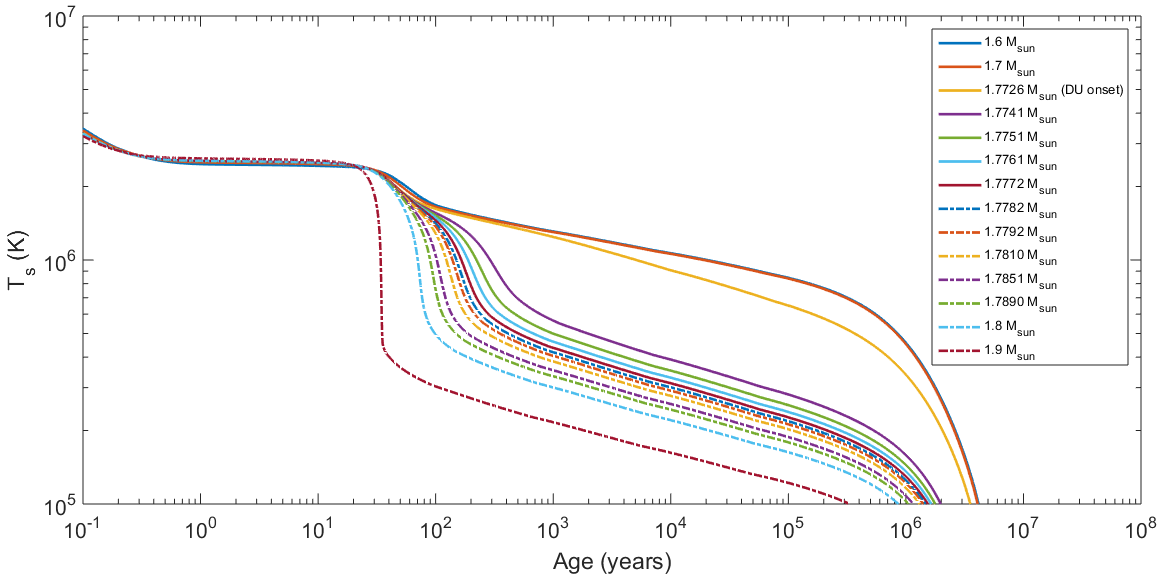}
\caption{Same as Fig.~\ref{bsr8cool}, but for model IU-FSU.}
\label{iufsu}
\end{figure}

The cooling of a wide range of masses for each model studied  is shown in Figs.~\ref{bsr8cool}-\ref{iufsu}.
As we can see, all models exhibit, qualitatively, the same behavior, with lighter stars displaying slow cooling whereas heavier ones show fast cooling. Each model has a different mass at which the DU processes sets in, this can be traced back to the differences in the microscopic model, particularly to the symmetry energy, its slope and curvature. We must note that model IU-FSU sets itself a part due to a lower symmetry energy slope and much higher curvature. This leads to the DU onset to take place at stars with much higher masses. Regardless of the microscopic model, or the mass at which the DU sets in, one can see a substantial difference in the cooling curves once it sets in. The reason behind such behavior lies in the strength of the DU process ($\sim 10^{27} (T_9)^6$ erg/cm$^3$s, being $T_9$ the temperature in units of $10^9 K$), much higher than that of the modified Urca process ($\sim 10^{21} (T_9)^6$ erg/cm$^3$s). Thus even if active in just a small kernel at the stellar core, it strongly affects the thermal evolution of star, as shown in Figs.~\ref{bsr8cool}-\ref{iufsu}.

One also sees that as the star's mass increase the cooling becomes faster -- which can be explained by the fact that the DU kernel at the star's center is growing with the mass of the star. Eventually the DU kernel becomes large enough that any increase in its size becomes mostly irrelevant and the cooling behavior of the star changes very little with any increase in the mass. 

\begin{figure}[h!]
\centering
\includegraphics[width = 9 cm, height = 5 cm]{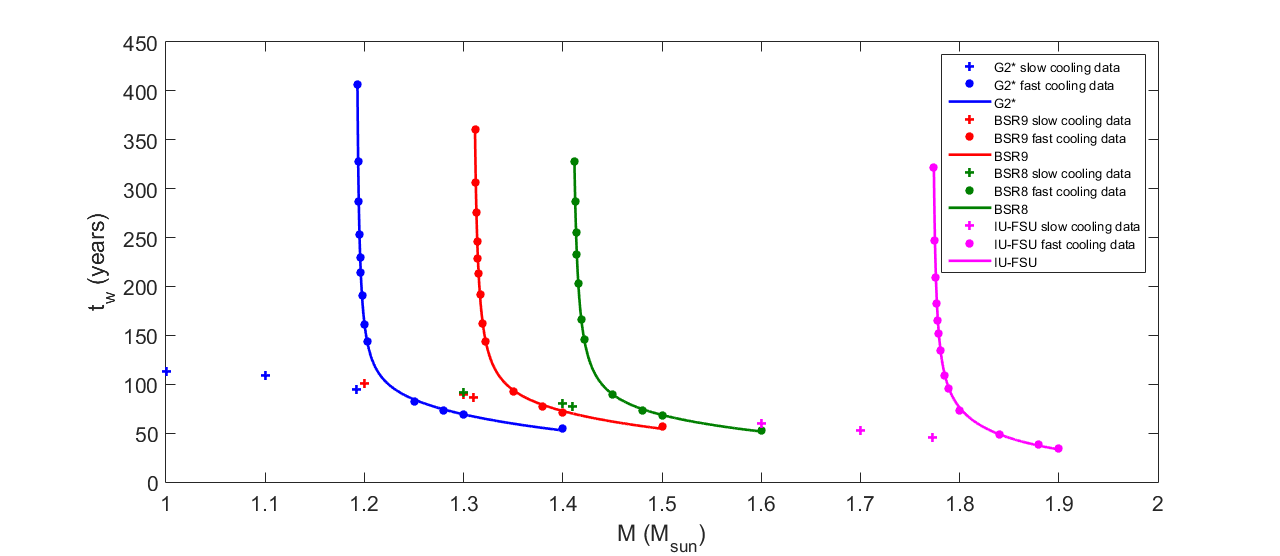}
\caption{Relaxation time as a function of gravitational mass for the different models studied. The discontinuity indicates the DU onset for each model. We note that the solid line is a fit for non-linear region. We omit the fit for the linear part as not to overload the figure.}
\label{tw_x_M}
\end{figure}

We now use the definition given by eq.~(\ref{tw}), to determine the thermal relaxation time of the stars whose cooling is depicted in Figs.~\ref{bsr8cool}-\ref{iufsu}. The results are shown in Fig.~\ref{tw_x_M}.

One can see that the thermal relaxation time  exhibits a highly non-linear behavior, drastically increasing at certain values of mass and quickly decreasing to a smoother shape. The mass at which the relaxation time increases is precisely the mass at which the DU sets in (for that particular microscopic model). The overall decrease of the thermal relaxation time with the increase of the mass has already been identified in previous studies, however, in order to identify the non-linear behavior near the DU onset, one needs a high-resolution study of the cooling of stars just above such onset, as in these stars the DU process is not pervasive in the star's interior as we will discuss below.

\begin{figure}[h!]
\centering
\includegraphics[width =9 cm, height = 5 cm]{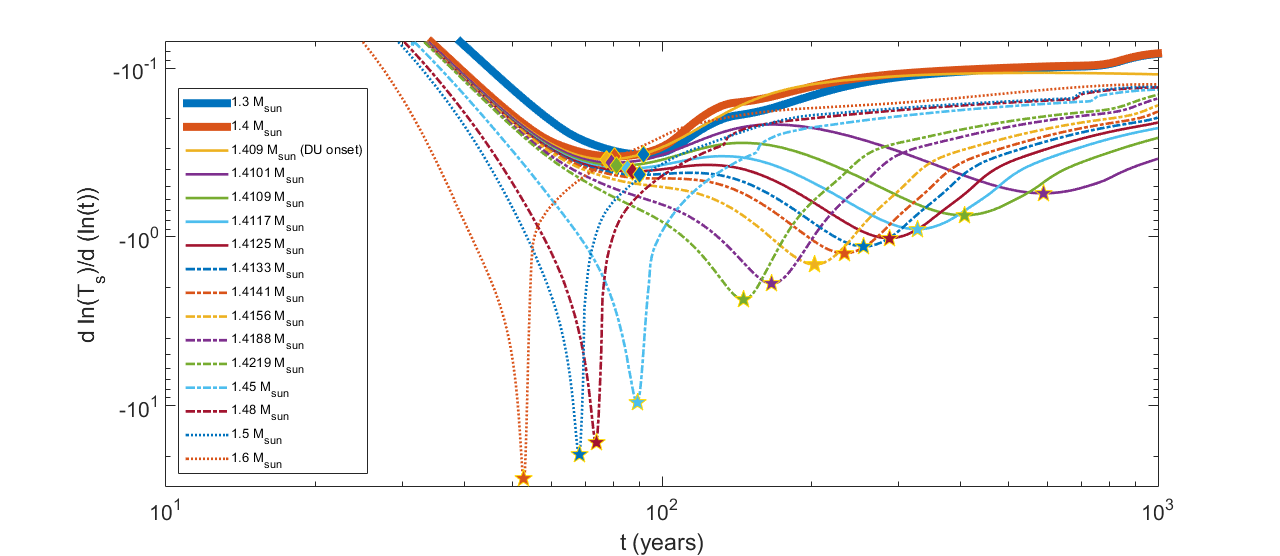}
\caption{$d\ln(T_s)/d(\ln(t))$ as a function of time for the thermal evolution of stars described by the BSR8 model. The minimum of each curve represents the thermal relaxation time.}
\label{dlnt_bsr8}
\end{figure}

\begin{figure}[h!]
\centering
\includegraphics[width =9 cm, height = 5 cm]{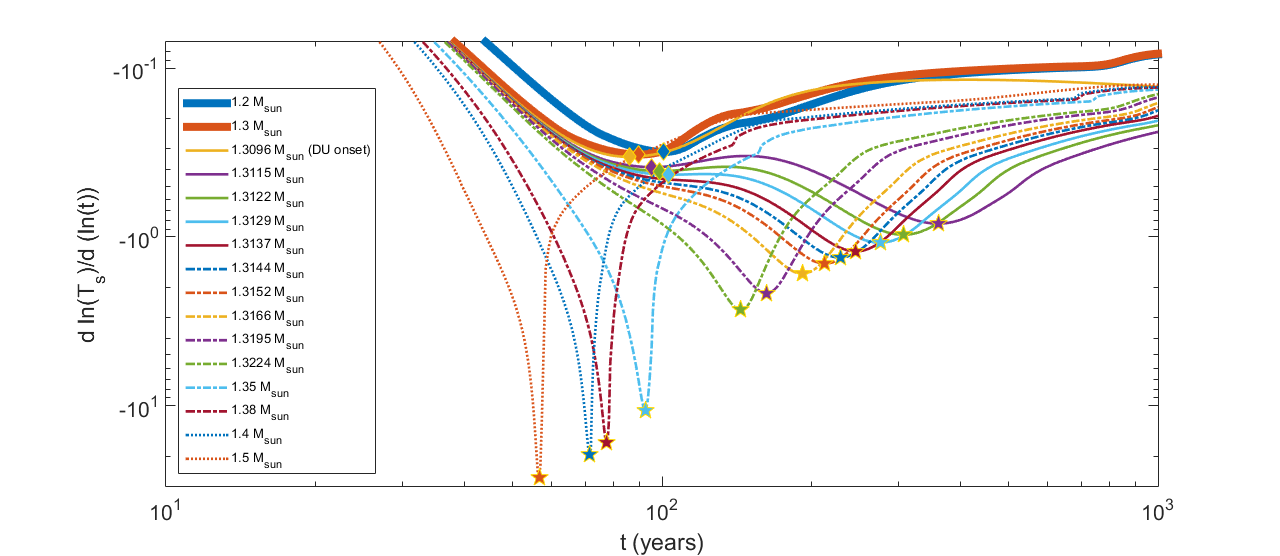}
\caption{Same as Fig.~\ref{dlnt_bsr8}, but for model BSR9.}
\label{dlnt_bsr9}
\end{figure}

\begin{figure}[h!]
\centering
\includegraphics[width = 9 cm, height = 5 cm]{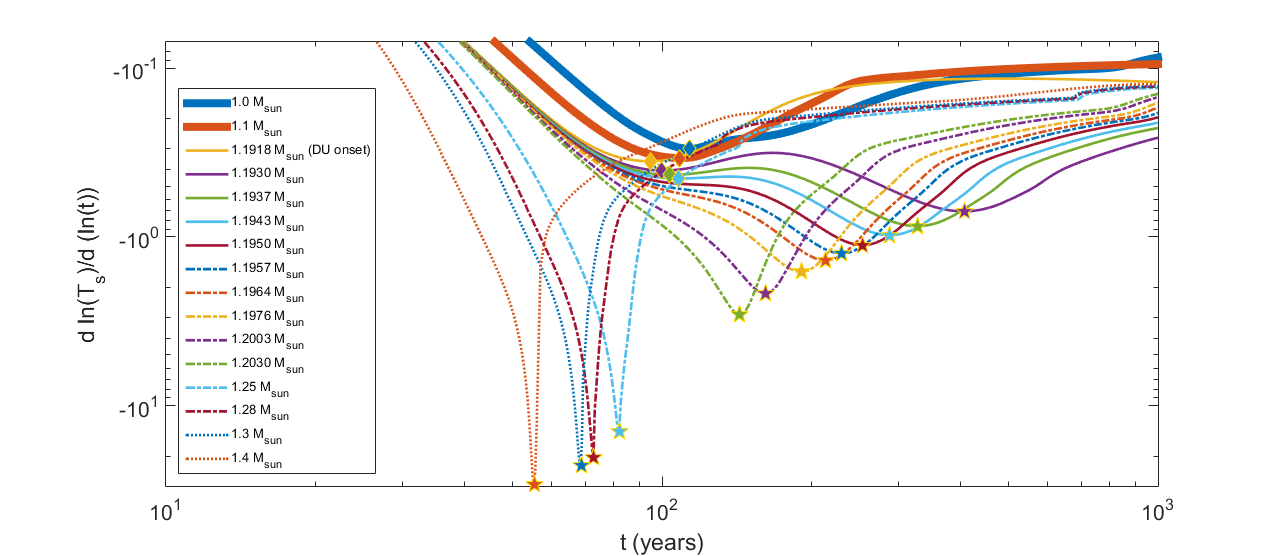}
\caption{Same as Fig.~\ref{dlnt_bsr8}, but for model G2*.}
\label{dlnt_g2star}
\end{figure}

\begin{figure}[h!]
\centering
\includegraphics[width = 9 cm, height = 5 cm]{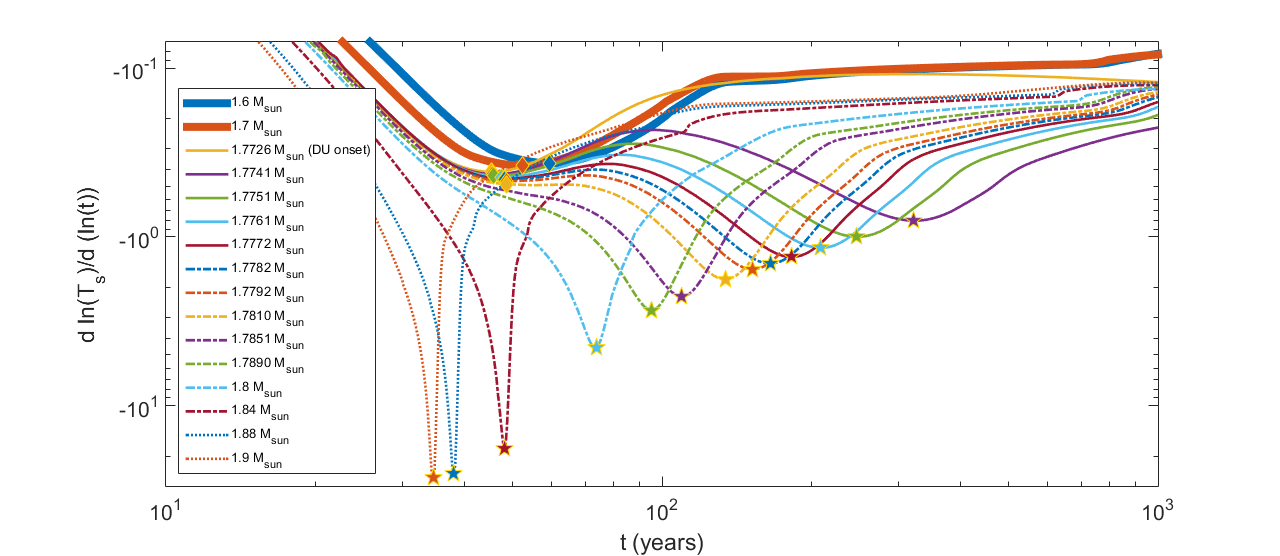}
\caption{Same as Fig.~\ref{dlnt_bsr8}, but for model IU-FSU.}
\label{dlnt_iufsu}
\end{figure}

To better understand the non-linear behavior of $t_w$ it is helpful to analyze the evolution of the derivative of $\ln T_s$, which we show in Figs.~\ref{dlnt_bsr8}--\ref{dlnt_iufsu}.

\begin{figure}[h!]
\centering
\includegraphics[width =9 cm, height = 5 cm]{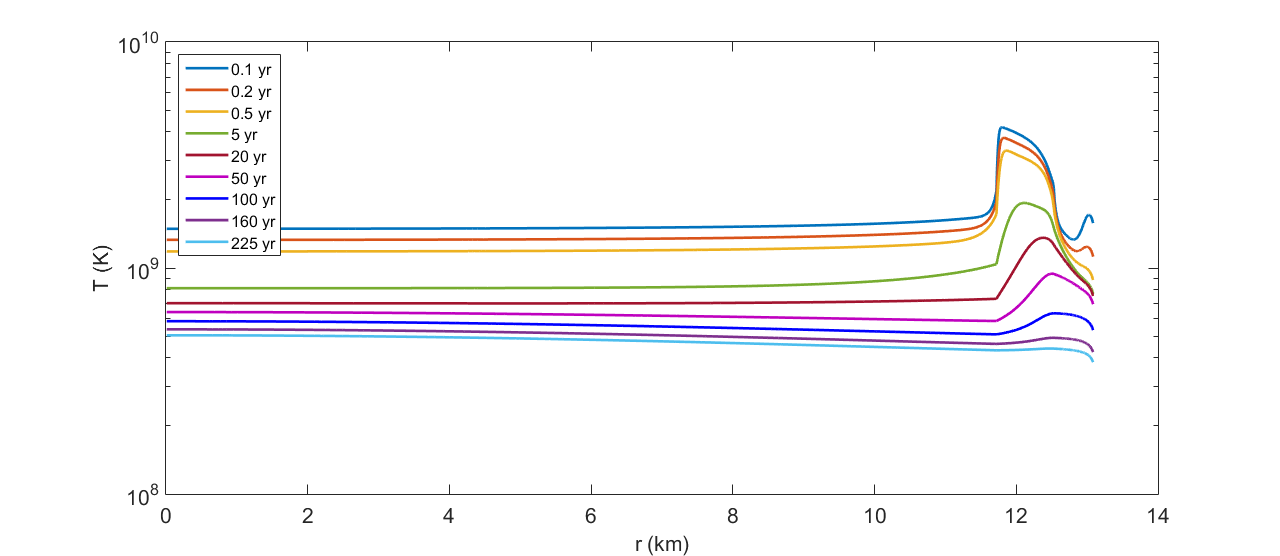}
\
\caption{Temperature profile inside a 1.3 solar mass star within the BSR8 model. Each curve represents different snapshots with the age indicated in the legends.}
\label{1.3prof}
\end{figure}

\begin{figure}[h!]
\centering
\includegraphics[width =9 cm, height = 5 cm]{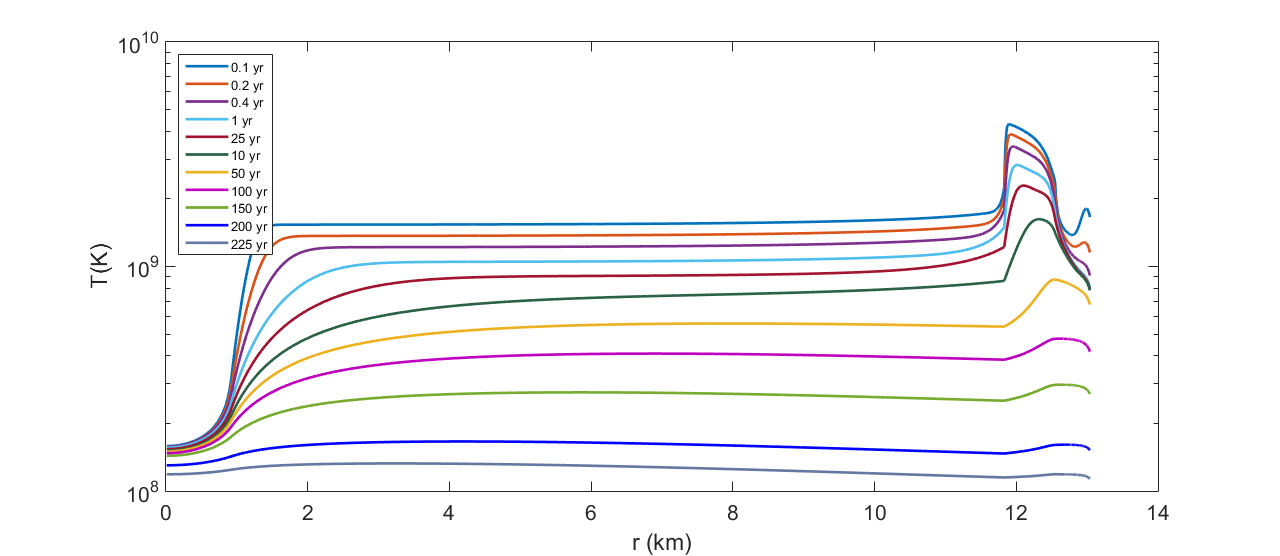}
\caption{Same as fig.\ref{1.3prof} but for a 1.41 $M_{\odot}$ star.}
\label{1.41prof}
\end{figure}

\begin{figure}[h!]
\centering
\includegraphics[width =9 cm, height = 5 cm]{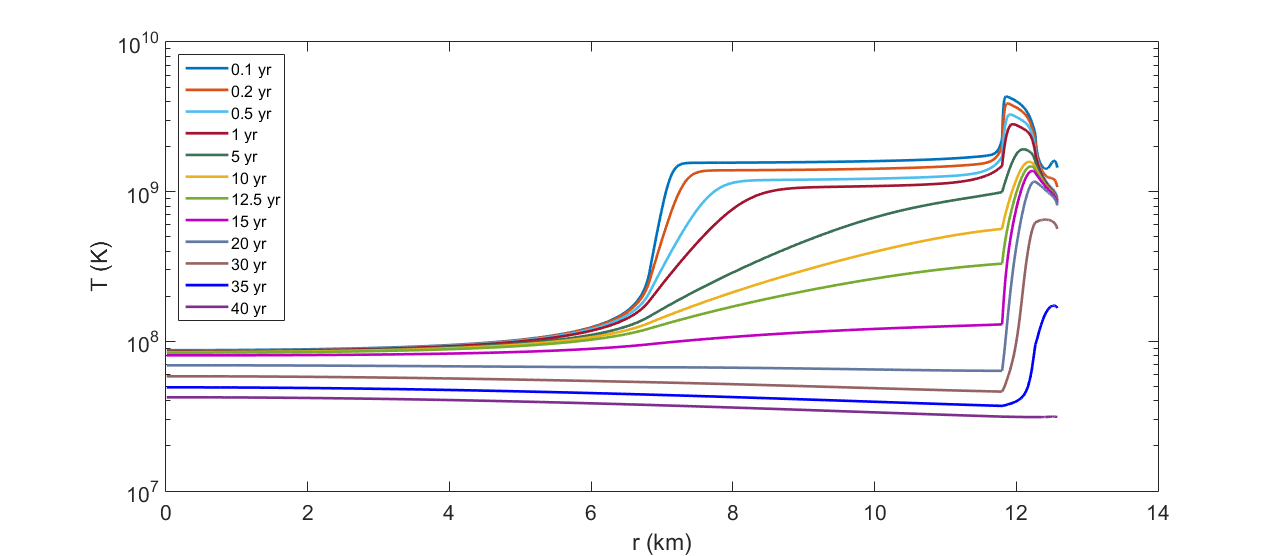}
\caption{Same as fig.\ref{1.3prof} but for a 1.8 $M_{\odot}$ star.}
\label{1.8prof}
\end{figure}

One can see that for stars with slow cooling, the relaxation process is smoother, as can be seen by the smooth minimum in the curves of Figs.~\ref{dlnt_bsr8}--\ref{dlnt_iufsu}, located at  $\sim 100$ years.
%For stars with fast cooling (due to prominent DU process), the minimum is significantly sharper, as also seen in the graphs.
Analyzing Figs.~\ref{dlnt_bsr8}--\ref{dlnt_iufsu} we also see that as the stellar mass increases and the DU process is triggered, a second minimum appears at later times. This minimum has a larger magnitude than the first as it is associated with the DU process. It is also clear that as the mass increases (and so does the region in which the DU takes place) this second minimum becomes sharper and more intense, until eventually overtaking the first smoother minimum - leading to stars with one minimum only, except this time much deeper. 
This indicates that stars with small DU kernels in the interiors have a ``double'' thermalization process. Such stars exhibit the thermal behavior typical of stars with and without the DU process. In order to understand this process we show the temperature profile at different times for stars of different masses (all in the BSR8 model): i) $ M = 1.3 M_\odot$ (no DU PROCESS, fig.~\ref{1.3prof}); ii)  $ M = 1.41 M_\odot$ (just above the onset of the DU, fig.~\ref{1.41prof}); and iii) $M = 1.8 M_\odot$ (well above the DU onset -- prominent DU, fig.~\ref{1.8prof}). We note that all other models exhibit, qualitatively, the same behavior and we omit the figures for the sake of conciseness.

A careful examination of Figs.~\ref{1.3prof} - \ref{1.8prof} provides important insight on the behavior discussed above. First we  review the thermal evolution of a low mass star, such as that exhibited in Fig.~\ref{1.3prof}. In such stars the DU process is absent, thus there is no fast cooling mechanism within. This means that the core cools down in a mostly uniform matter, and as such by 0.1 years it is mostly isothermal. It is also relatively hot. Thus, in stars such as this, the thermal relaxation is solely due to the thermal coupling between the core and the crust. Due to the relatively high temperature of the core this process is slow and smooth. Now we shall skip to high mass stars, such as the 1.8 $M_\odot$, whose temperature profiles are exhibited in Fig.~\ref{1.8prof}. In these stars the DU process takes place in a large region of their core (although not the entire core). As such, this leads to a temperature profile with a large temperature gradient between the DU and non-DU region within the core. The DU region is significantly colder and acts as a very strong heat sink due to its size and to the strength of the DU process. As such when the core becomes isothermal, it does so at a lower temperature when compared to the non-DU cooling. Furthermore, the core acts as a stronger heat sink, efficiently drawing heat from the crust, thus exhibiting the behavior of a strong cold front, that readily reaches the surface - leading to a sudden and strong drop in surface temperature. Finally, we now turn our attention to stars just above the onset of the DU - illustrated by the $ M=1.41 M_\odot$ star - Fig.~\ref{1.41prof}. These stars have the DU process limited to a small kernel in their interior. This can be seen in Fig.~\ref{1.41prof} as the relatively small region with a lower temperature at small radius. Due to the small size of this region, its influence on the global thermal behavior of the star is limited. As such, for the initial years, the outer core and the crust of the star behave as if there is no DU process and the star (initially) behave as a slow cooling star. Eventually the core will become isothermal, as the influence of the DU region propagates until it eventually reaches the surface, leading to a belated thermal relaxation. This indicates a hybrid behavior for such stars, in which at young ages it behaves as a non-DU stars and later exhibits a drop in its surface temperature. The result of which is an abnormally high relaxation time for stars in this transitional region. 
This is an exotic behavior: stars near the onset of the DU process may exhibit larger thermal relaxation times than other objects either with or without prominent fast cooling processes.

The scenario discussed above can be further understood by analyzing the conductive luminosity \cite{1999Weber..book} within the star. This quantity is related to a fraction of the total energy of the star that is transferred via heat conduction within the star. For a general relativistic star with spherical symmetry it can be defined as
\begin{equation}
    L_r = -4 \pi r^2 \kappa(r) \sqrt{1-\frac{m(r)}{r}}e^{-\phi}\frac{d}{dr}(T e^\phi).
\end{equation}

\begin{figure}[h!]
\centering
\includegraphics[width =9 cm, height = 5 cm]{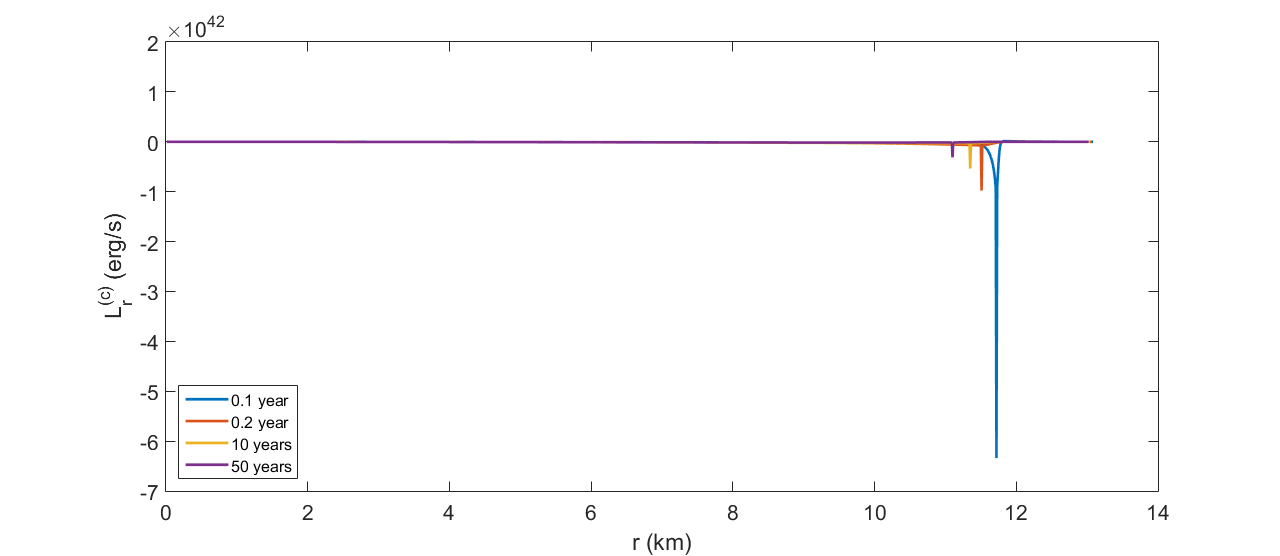}
\caption{Conductive luminosity for the 1.3 solar mass star of the BSR8 model at different times. The large peaks indicate the large temperature gradients between the core and the crust.}
\label{Lcr_13}
\end{figure}

\begin{figure}[h!]
\centering
\includegraphics[width =9 cm, height = 5 cm]{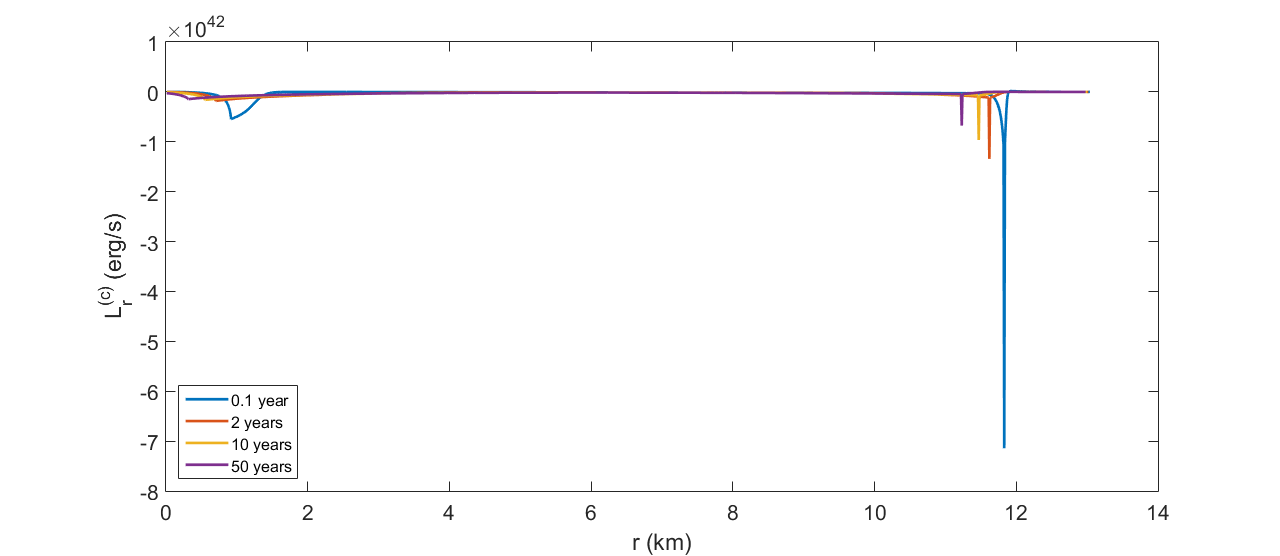}
\caption{Same as Fig.\ref{Lcr_13} but for a 1.41 $M_\odot$ star (just above the DU onset). The large peaks at large radius represent the temperature gradient between core and crust. One also notice a small non-zero conductivity at small radii. This region is associated with the small DU kernel at the star's core.}
\label{Lcr_14156}
\end{figure}

\begin{figure}[h!]
\centering
\includegraphics[width =9 cm, height = 5 cm]{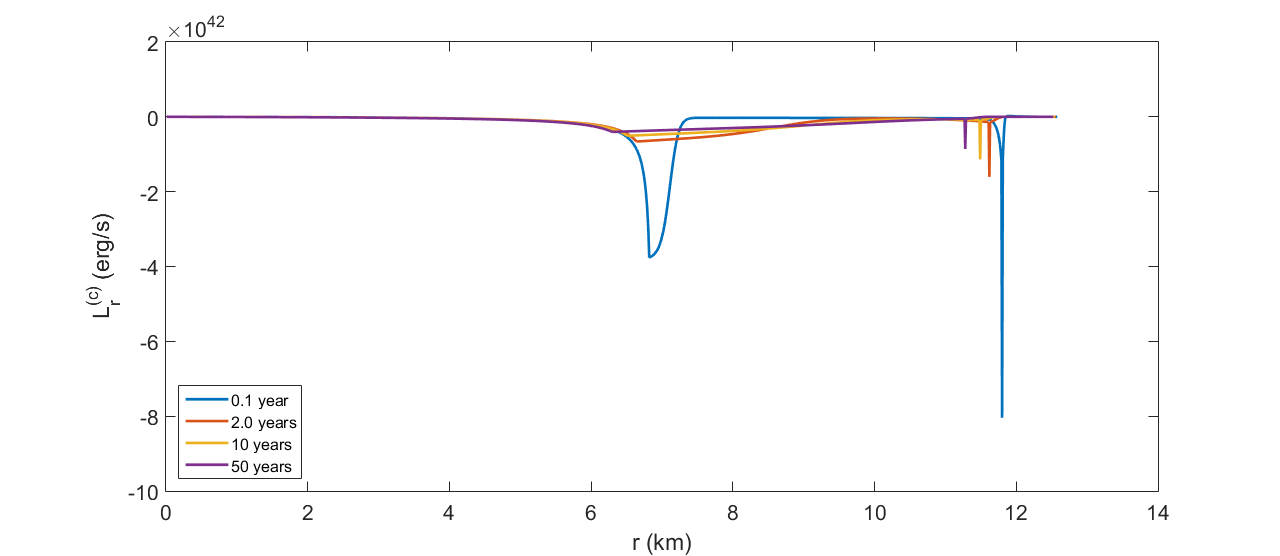}
\caption{Same as fig.\ref{Lcr_13} but for a 1.8 $M_\odot$ star - well above the DU onset. Besides the large peaks representing the core-crust temperature gradient we also have a large peak representing the temperature gradient between the DU and non-DU regions of the star.}
\label{Lcr_18}
\end{figure}

We show in Figs~\ref{Lcr_13}-\ref{Lcr_18}  the conductive luminosity for the three stars discussed above.

The results of Figs~\ref{Lcr_13}-\ref{Lcr_18} corroborate our previous assessment. We see in Fig.~\ref{Lcr_13} that the core is mostly isothermal by 0.1 year (as indicated by a mostly absent  conductivity luminosity within the core), leading to very little heat transport within the core - being the core-crust interface the most prominent conductive heat sink (with the core absorbing large amounts of the crust heat). Fig.~\ref{Lcr_13} also shows that with the passage of time this sink becomes smaller until core-crust thermal equilibrium is achieved. The situation is drastically different for the 1.8 solar mass star, in which the DU process is present in a large part of the core as shown in  Fig.~\ref{Lcr_18}. In this star we see that besides the core-crust heat sink, there is also a significant heat sink within the core - which is associated with the interface between the DU and non-DU regions (which is naturally smoother than the core-crust interface). In this system we have a strong heat sink reducing the temperature of the core, as well as the core absorbing heat from of the crust. Finally, in Fig.~\ref{Lcr_14156} we show the conductive luminosity for the 1.41 solar mass star - which is just above the DU onset. As discussed before, this star has a hybrid behavior. We see the expected peak at the core-crust transition and a relatively small heat sink near the star's core. This is associated to the small DU kernel present in this star. As such, this small heat sink takes longer to make itself noticeable and the star initially cools down as if there was no DU process - until the effect of this region finally reaches the crust leading to a second belated thermal coupling, indicated by the late drop in the temperature of the surface.

\begin{figure}[h!]
\centering
\includegraphics[width =9 cm, height = 5 cm]{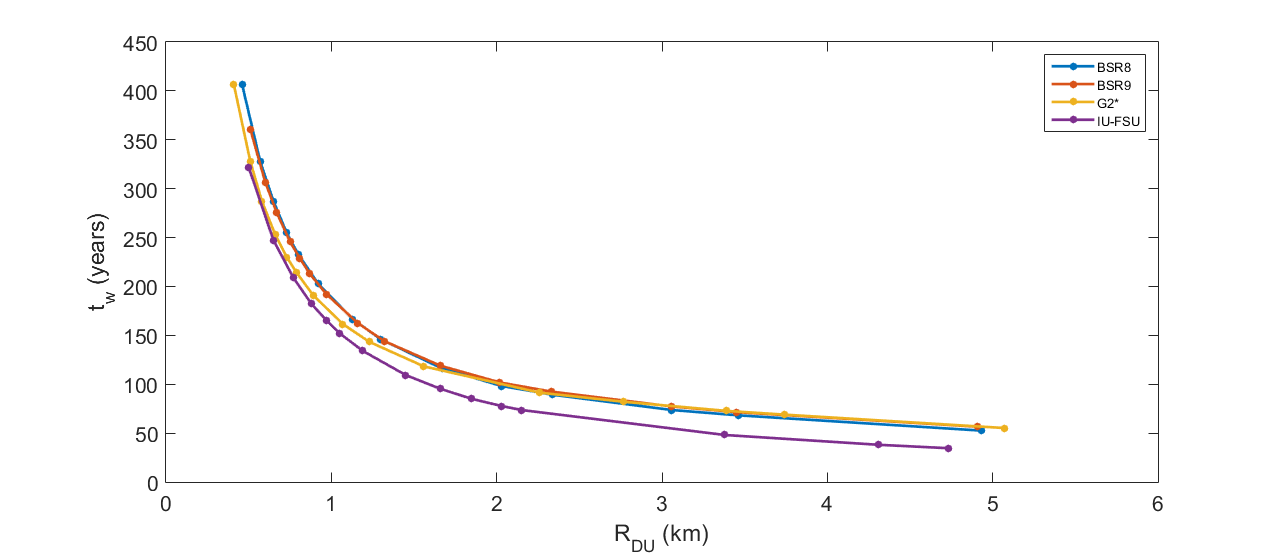}
\caption{Thermal relaxation time ($t_w$) as a function of direct Urca (DU) kernel's radius - for stars with active DU in their interior.}
\label{tw_Rdu}
\end{figure}

Finally, we discuss what happens as the mass of the star increases and we move from this transitional behavior to the well known fast cooling scenario. As discussed above, after a certain point, when the DU is pervasive enough the thermal relaxation time changes very little with the increase of the mass. This can be clearly seen in Fig.~\ref{tw_x_M}. In order to understand how large the DU process kernel needs to be such that the star finds itself out of the transient region, we analyse the dependence of the relaxation time on radius of the DU kernel (the fraction of the core in which the DU takes place). In Fig.~\ref{tw_Rdu} we show how the thermal relaxation time changes as the radius of the DU kernel ($R_{DU}$) increases (for stars of higher masses). One can note that this graph is essentially analogous as that shown in Fig.~\ref{tw_x_M}, as the DU kernel increases together with the mass. It is, nonetheless, useful to see the direct dependence on the size of the DU kernel, as it tell us that the thermal relaxation time stabilizes at $R_{DU} \sim 2 -3$ km, meaning that at this point the star will behave as expected for a fast cooling object.

\section{Conclusions}

In this work we revisited the concept of relaxation time for the cooling of neutron stars -- we used a set of equations of state to investigate how the thermal relaxation time depends on the micro and macroscopic properties of the star. Previous studies \cite{Gnedin2001,Lattimer1994} have found a direct dependence between relaxation time and properties such as crust thickness, stellar mass and radius; in fact it depends linearly on a quantity denoted $\alpha$, which in turn depends only on these stellar macroscopic properties -- the proportionality constant in such dependence is given by $t_1$ (see eq.~(\ref{tw})) and is associated with the microscopic properties of the model adopted. Our study confirms these results, while adding a new attribute: a nonlinear increase on the relaxation time for stars near the onset of the direct Urca process. We have found that stars with masses just above that of the onset of the DU process will have longer relaxation times. In order to understand this behavior we performed an extensive analysis of the relaxation time of stars just above the onset of the DU process. We have found (across all models) that as the DU process sets in a second minimum appears at later times in the curve of $d\ln(T_s)/d(\ln(t))$. We have also found that as the mass increases such minimum becomes stronger and happens at earlier times. We have identified this second minimum as a late influence of the DU process happening in small regions at the core of the stars. This is confirmed by the analysis of both the temperature distribution and conductive luminosity inside the stars. We can see that for stars with low mas, the core is mostly isothermal and acts as a heat sink, drawing heat from the crust. As for high mass stars with prominent DU process in their interior, we have colder region of the core (where the DU is active) that draws heat from  the non-DU region, whereas this region draws heat from the crust. This causes a quick and strong cold-front that leads to the well known drop in temperature surface for fast cooling stars. For stars in the transition from absent to prominent DU - we have a hybrid behavior: due to the smaller regions in which the DU is active, the star initially behaves as if there were no DU, with a smooth coupling between core and crust, which is followed by the late arrival of the DU influence, leading to a belated thermal relaxation time. Our studies show that this is a transitional phenomenon, taking place in stars whose composition is just above the onset of the DU process. We have determined that as the DU kernel in the star grows, the object returns to the expected behavior, with shorter relaxation times for stars of higher masses. Our study indicates that, for direct Urca kernels reaching a size of $\sim 2-3$ km, the thermal relaxation time regains its normal and expected behavior.

For this study we have used several microscopic models, covering a wide range of microscopic properties and leading to stars with different masses at the onset of the DU process. All models exhibited the same qualitative behavior, which leads us to believe this is not model dependent. As mentioned before, however, we have not taken into account pairing - which reduces considerably the  neutrino emissivity strength, and may even lead to the total suppression of the DU process. We have opted to leave it out of this study as the current uncertainties on pairing at high density regimes would obscure our analysis. We stress that we are not claiming pairing is irrelevant, only that we left it out for the purpose of qualifying the phenomenon studied here. We intend to pursue further investigation of the phenomena we found here by taking account several models of superfluidity, accounting for possible proton and neutron pairings covering different regions of the star. 
We believe, nonetheless, that we have found an interesting phenomena, possibly not seen before - one that allows neutron stars to exhibit unusually large relaxation time if their structure happens to have just reached the onset of fast neutrino emissivity, typically associated with the DU process. For future perspective we intend to extend this study (in addition to the aforementioned pairing) to other possible transitional phenomena, such as in hybrid stars. It may be possible that for a hybrid EoS that allows for stars in its family to possess a quark matter core to exhibit similar behavior, with a change in the relaxation time for stars that have just reached the transition to quark matter. Such studies are currently underway and will be discussed in future publications.

\section*{Acknowledgements}
T.S. and R.N. acknowledges financial support from Coorden\c c\~ao de Aperfei\c coamento de Pessoal de N\' ivel Superior (CAPES) and Conselho Nacional de Desenvolvimento Cient\'ifico e Tecnol\'ogico (CNPq). R.N. acknowledges financial support from CAPES, CNPq and FAPERJ. This work is part of the project INCT-FNA Proc. No. 464898/2014-5 as well as FAPERJ JCNE Proc. No. E-26/203.299/2017. This work is also partially supported by CNPq under grants 310242/2017-7 and 406958/2018-1 (O.L.), 433369/2018-3 (M.D.), and by Funda\c{c}\~ao de Amparo \`a Pesquisa do Estado de S\~ao Paulo (FAPESP) under the thematic project 2013/26258-4 (O.L.) and 2017/05660-0 (O.L., M.D.).

\bibliography{thermal_relaxation}
\end{document}